\documentclass[prl,twocolumn,showpacs,preprintnumbers,superscriptaddress]{revtex4}%
\usepackage{graphicx}
\usepackage{dcolumn}
\usepackage{bm,hyperref}
\usepackage{amsmath}
\usepackage{amsfonts}
\usepackage{amssymb}%
\providecommand{\U}[1]{\protect\rule{.1in}{.1in}}
\begin{document}

\title{Asymmetric transmission of linearly polarized light at optical metamaterials}
\author{C. Menzel}
\affiliation{Institute of Condensed Matter Theory and Solid State
Optics, Friedrich-Schiller-Universit\"at Jena, Max-Wien-Platz 1,
D-07743 Jena, Germany}

\author{C. Helgert}
\affiliation{ Institute of Applied Physics, Friedrich-Schiller-Universit\"at Jena,
Max-Wien-Platz 1, 07743 Jena, Germany}

\author{C. Rockstuhl}
\affiliation{Institute of Condensed Matter Theory and Solid State
Optics, Friedrich-Schiller-Universit\"at Jena, Max-Wien-Platz 1,
D-07743 Jena, Germany}

\author{E.-B. Kley}
\affiliation{ Institute of Applied Physics, Friedrich-Schiller-Universit\"at Jena,
Max-Wien-Platz 1, 07743 Jena, Germany}

\author{A. T\"unnermann}
\affiliation{ Institute of Applied Physics, Friedrich-Schiller-Universit\"at Jena,
Max-Wien-Platz 1, 07743 Jena, Germany}

\author{T. Pertsch}
\affiliation{ Institute of Applied Physics, Friedrich-Schiller-Universit\"at Jena,
Max-Wien-Platz 1, 07743 Jena, Germany}

\author{F. Lederer}
\affiliation{Institute of Condensed Matter Theory and Solid State
Optics, Friedrich-Schiller-Universit\"at Jena, Max-Wien-Platz 1,
D-07743 Jena, Germany}

\begin{abstract}
We experimentally demonstrate  a three-dimensional chiral optical metamaterial
that exhibits an asymmetric transmission for forwardly and backwardly
propagating linearly polarized light. The observation of this novel effect requires
a metamaterial composed of three-dimensional chiral metaatoms without any
rotational symmetry.  Our analysis is supported by a systematic investigation of
the transmission matrices for arbitrarily complex, lossy media that allows
deriving a simple criterion for asymmetric transmission in an arbitrary
polarization base. Contrary to physical intuition,  in general the polarization
eigenstates in such three-dimensional and low-symmetry metamaterials do not
obey fixed relations and the associated transmission matrices cannot be
symmetrized.
\end{abstract}

\pacs{XX.XX.XX} \maketitle

During the past several years optical metamaterials (MMs) have attracted an
enormous interest since they promise to allow for a manipulation of light
propagation to a seemingly arbitrary extent. MMs are usually obtained by
assembling sub-wavelength unit cell structures called metaatoms. Initial studies
on MMs were based on  rather simple and highly symmetric metaatoms
\cite{Soukoulis2007,Zhang2005,Kafesaki2007}. Recently, more and more
sophisticated structures were explored in order to achieve customized
functionalities like, e.g. a negative refractive index due to chirality
\cite{Zhang2009,Plum2009,Wang2009}. Also, a large variety of plasmonic
metaatoms were investigated that evoke a huge polarization rotation like
gammadions, omega shaped particles or helices
\cite{Wang2009b,Rockstuhl2009,Gansel2009}. Studying the characteristics of
light propagation in such low-symmetry MMs also revealed unexpected
phenomena like asymmetric transmission for circularly polarized light
\cite{Fedotov2006, Schwanecke2008, Drezet2008a}. Although at first sight this
effect of nonreciprocal transmission, to date not observed for linearly polarized
light, is counterintuitive, it does not violate Lorentz' reciprocity theorem. This
asymmetric transmission of circularly polarized light was demonstrated at
so-called planar chiral MMs. Such MMs are composed of metaatoms without
structural variation in the principal propagation direction. They preserve
symmetry in this direction and are only chiral in the two-dimensional space
\cite{Arnaut1997}; thus, strictly speaking, they are intrinsically achiral in three
dimensions since the mirror image of a structure is congruent with the structure
itself if operated from the backside. The remaining mirroring plane is
perpendicular to the propagation direction.

In this Letter we theoretically and experimentally demonstrate a novel MM
design which breaks the latter symmetry. For the first time our approach reveals
that the very structures exhibit asymmetric transmission for linearly polarized
light. We emphasize that also in this case the reciprocity theorem is not violated
since only reciprocal materials are involved.

Prior to any further considerations we concisely discuss the effect of a potential
MM substrate that is, after all, in most cases required for fabricating planar
MMs. Generally speaking, just this supporting substrate breaks the mirror
symmetry for any planar structure perpendicular to the propagation direction
\cite{Maslovski2009}. However, the effect of this symmetry
 breaking  is almost negligible when compared to the impact of a strong
structural anisotropy that leads e.g. to asymmetric transmission of circularly
polarized light \cite{Fedotov2006, Schwanecke2008}. On the other hand, if no
structural anisotropy is present, like e.g., for gammadions, the chirality of the
planar metatom-substrate system yields a measurable polarization rotation
\cite{Menzel2008}. So, although asymmetric transmission for linearly polarized
light may have been observed already, its magnitude was too weak and it was
merely attributed to an insufficient measurement accuracy rather than to a true
physical mechanism. To unambiguously enhance the effect towards a
measureable extent, it is necessary to deliberately break the symmetry in the
third dimension of the metaatom itself and not just relying on the rather
perturbative effect of a supporting substrate. Exactly this goal is pursued in the
present work.

We start in deriving a criterion for the occurrence of asymmetric transmission in
an arbitrary polarization base and particularly for linearly polarized light. The
transmission of coherent light through any dispersive optical system can be
described by means of complex Jones matrices T \cite{Fedotov2006,
Drezet2008a}. We consider an incoming plane wave that propagates in positive
$z$-direction
$$\mathbf{E}_\mathrm{i}(\mathbf{r},t)=\begin{pmatrix}I_x \\ I_y\end{pmatrix}e^{i(kz-\omega t)},$$
with frequency $\omega$, wave vector $k$ and complex amplitudes $I_x$ and
$I_y$. The transmitted field is then given by
$$\mathbf{E}_\mathrm{t}(\mathbf{r},t)=\begin{pmatrix}T_x \\ T_y\end{pmatrix}e^{i(kz-\omega t)},$$
where it is assumed that the medium is symmetrically embedded e.g. in vacuum.
The T - matrix relates the generally complex amplitudes of the incident field to
the complex amplitudes of the transmitted field:

\begin{equation}
\begin{pmatrix}T_x \\ T_y\end{pmatrix}=\begin{pmatrix}T_{xx} & T_{xy} \\ T_{yx} & T_{yy} \end{pmatrix}\begin{pmatrix}I_x
\\ I_y\end{pmatrix}=\begin{pmatrix}A & B \\ C & D \end{pmatrix}\begin{pmatrix}
I_x \\ I_y\end{pmatrix}=\hat{T}^\mathrm{f}_{\mathrm{lin}}\begin{pmatrix}I_x \\ I_y\end{pmatrix},\label{EQ_LinForward}
\end{equation}
where we have replaced  $T_{ij}$ by $A,B,C,D$ for convenience. The indices $\mathrm{f}$
and $\mathrm{lin}$ indicate propagation in forward direction and a special linear base with
base vectors parallel to the coordinate axes, i.e. decomposing the field into $x$-
and $y$-polarized light. Applying the reciprocity theorem delivers the T-matrix
$\hat{T}^\mathrm{b}$ for  propagation in $-z$ or backward direction

\begin{equation}
\hat{T}^\mathrm{b}_{\mathrm{lin}}=\begin{pmatrix}A & -C \\ -B & D \end{pmatrix}\label{EQ_LinBackward}
\end{equation}
characterizing the transmission in a fixed coordinate system with the sample
being rotated by $180^{\circ}$ with respect to either $x-$ or $y-$axis. The form
of Eq.~\ref{EQ_LinBackward} results from the reciprocity of four-port systems
and the definition of the coordinate system \cite{Potton2004}. In general, all components
$A,B,C$ and $D$ are mutually different and strongly dispersive. Only for certain
symmetries they obey fixed relations.
For example for planar, anisotropic 2D chiral metaatoms the mirror image
coincides with that seen from the backside, i.e. $B=C$ and therefore

\begin{equation}
\hat{T}^\mathrm{f}_{\mathrm{lin}}=\begin{pmatrix}A & B \\ B & D \end{pmatrix}\hspace{2mm}
\textrm{and}\hspace{2mm}\hat{T}^\mathrm{b}_{\mathrm{lin}}=\begin{pmatrix}A & -B \\ -B & D \end{pmatrix}.
\label{EQ_Tmatrix_2D_chiral}
\end{equation}

On the other hand for gammadions on a substrate (or trapezoidally shaped) with
a $C_4$-symmetry with respect to the $z-$axis we have $D=A$ and $C=-B$ and
therefore

\begin{equation}
\hat{T}^\mathrm{f}_{\mathrm{lin}}=\hat{T}^\mathrm{b}_{\mathrm{lin}}=\begin{pmatrix}A & B \\ -B & A \end{pmatrix}.
\label{EQ_Tmatrix_C4_Chiral}
\end{equation}

For analytical as well as for experimental purposes it is advantageous to have at
hand the T-matrix also for circularly polarized light . It can be obtained by a
change of the base vectors from linear to circular states resulting in

\begin{equation}
\hat{T}^{\mathrm{f}}_{\mathrm{circ}}=\frac{1}{2}\begin{pmatrix}\left[A+D+i(B-C)\right] &
\left[A-D-i(B+C)\right] \\ \left[A-D+i(B+C)\right] &
\left[A+D-i(B-C)\right] \end{pmatrix},\label{EQ_T_CIRC_F}
\end{equation}
connecting the amplitudes of circularly polarized incident light with those of
circularly polarized transmitted light:
$$\begin{pmatrix}T_{+} \\ T_{-}\end{pmatrix}=T^{\mathrm{f}}_{\mathrm{circ}}\begin{pmatrix}I_{+} \\ I_{-}\end{pmatrix}=\begin{pmatrix}T_{++} & T_{+-} \\ T_{-+} &
T_{--}\end{pmatrix}\begin{pmatrix}I_{+} \\ I_{-}\end{pmatrix}.$$
by using Eq.~\ref{EQ_LinBackward} we find for propagation in $-z$-direction
$$T^{\mathrm{b}}_{\mathrm{circ}}=\begin{pmatrix}T_{++} & T_{-+} \\ T_{+-} & T_{--}\end{pmatrix}.$$

Now we introduce the overall transmission in an arbitrary base. The base
vectors denoted by $\mathbf{e}_1$ and $\mathbf{e}_2$ are assumed to be
normalized and linearly independent. The overall transmission may then be
written as
$$\mathbf{T}^{\mathrm{f,b}}=T_1^{\mathrm{f,b}}\mathbf{e}_1+T_2^{\mathrm{f,b}}\mathbf{e}_2=\hat{T}^{\mathrm{f,b}}\left\{I_1\mathbf{e}_1+I_2\mathbf{e}_2\right\}.$$
Asymmetric transmission for a given base vector $i\in\{1,2\}$ can then be
defined as the difference between the transmitted intensities for different
propagation directions as
\begin{equation}
\Delta(I_i=1;I_j=0)\doteq\Delta^{(i)}=|\mathbf{T}^\mathrm{f}|^2-|\mathbf{T}^\mathrm{b}|^2.
\label{EQ_AsymmTransArb}
\end{equation}

For the linear and the circular base we obtain

\begin{equation}
\Delta_{\mathrm{lin}}^{(x)}=|C|^2-|B|^2=-\Delta_{\mathrm{lin}}^{(y),}
\label{EQ_AsymmTransLin}
\end{equation}
and
\begin{equation}
\Delta_{\mathrm{circ}}^{(+)}=|T_{-+}|^2-|T_{+-}|^2=-\Delta_{\mathrm{circ}}^{(-)},
\label{EQ_AsymmTransCirc}
\end{equation}
respectively.

Obviously, the effect of asymmetric transmission depends
on the specific base. For both bases given here it is exclusively determined by the
difference between the off-diagonal elements of the respective T-matrices. For
structures obeying e.g. Eq.~\ref{EQ_Tmatrix_2D_chiral} we have
$\Delta_{\mathrm{lin}}=0$ whereas $\Delta_{\mathrm{circ}}\neq 0$. Note that for an arbitrary base
the diagonal elements of the T-matrix for different propagation directions are
not identical and will also contribute to asymmetric transmission.

\begin{figure}
\centering
\includegraphics[width=84mm,angle=0] {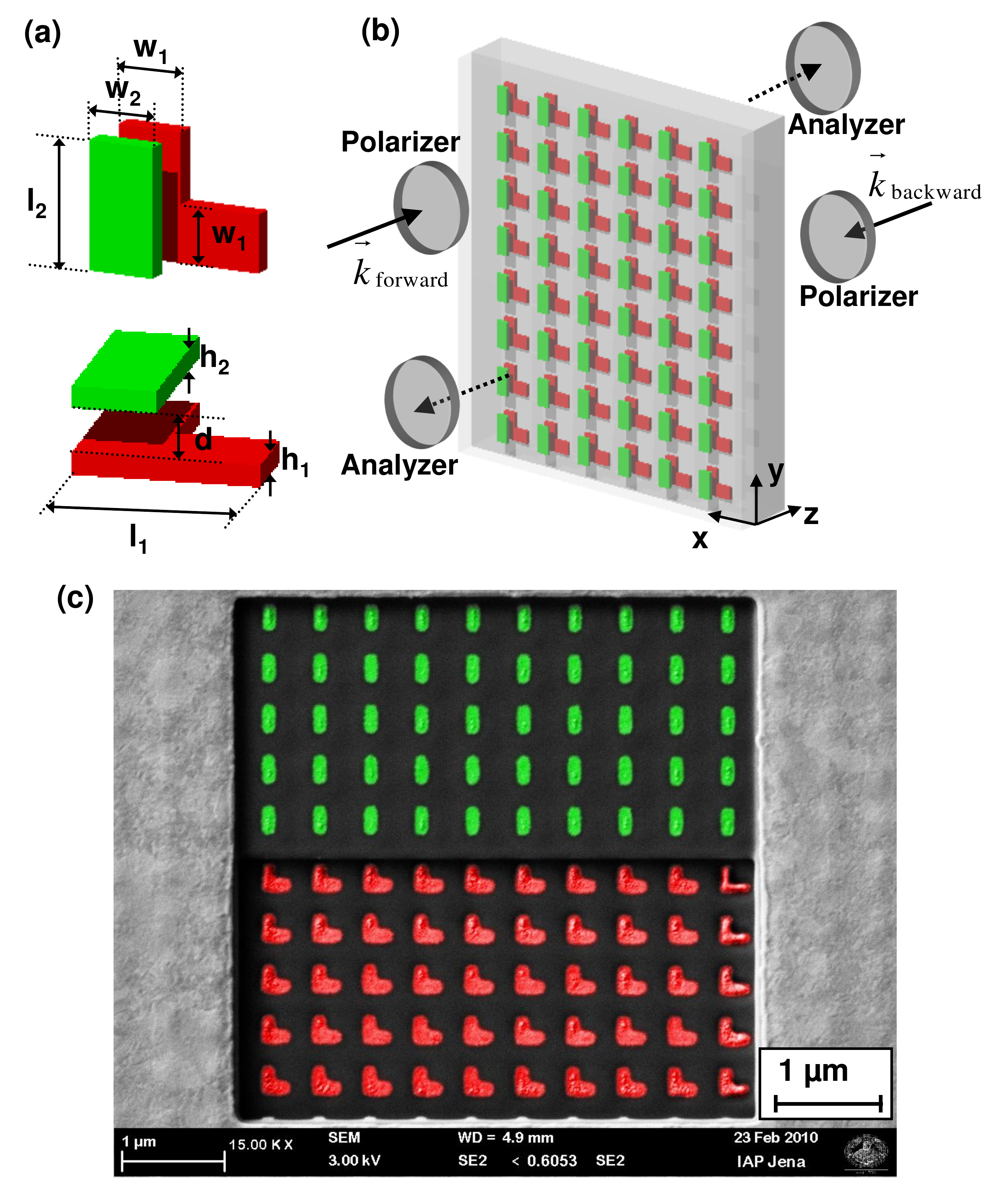}
\caption{a) Two sketches of the MM unit cell from different
perspectives with the definition of the geometrical parameters:
$l_1$=$l_2$=290\,nm, $w_1$=$w_2$=130\,nm, $h_1$=$h_2$=40\,nm,
$d$=80\,nm. b) Schematic of the experimental setup: Normally
incident light was linearly polarized before propagating through the
MM and subsequently analyzed by means of a second linear polarizer.
c) Normal view electron micrograph with false colors of the
fabricated MM. The meta-atoms were completely embedded in an
index-matched dielectric. Focused ion beam slicing reveals the two
layers composing the MM. Green and red colors represent the nanowires
and the L-structures in the top and bottom layer, respectively. The
periods in both the $x$ and $y$ directions are
500\,nm.}\label{FIG_1}
\end{figure}

From this analysis we can conclude that for MMs featuring $A\neq B\neq C\neq
D$ asymmetric transmission will be observable in any base and in particular for
linearly polarized light too. To prove this statement we designed a simple
metaatom that is essentially three-dimensional, non-symmetric and consists of
strongly coupled plasmonic elements. To rule out asymmetric effects due to the
presence of a substrate, the metaatoms are completely embedded in an
index-matched dielectric host. The geometry of the structure is shown in
Fig.~\ref{FIG_1}a. It consists of two closely spaced layers. The first layer
comprises an L-shaped metallic particle and the second one a single nanowire.
The L-shaped structure on its own causes an anisotropic response if both arms
are dissimilar, giving rise to asymmetric transmission for circularly polarized
light. Adding the nanowire breaks the remaining symmetry in the principal
propagation direction and leads to a genuine three-dimensional chiral metaatom.
Note the particles themselves, in general, sustain three first-order plasmonic
resonances; one may be associated with the nanowire and the two others with
the L-shaped particle. The coupling of these particles results in a complex
response featuring various resonances. The specific hybrid plasmonic
eigenmodes are of minor importance for our purpose and beyond the scope of
this work; though they can be easily accessed within the framework an oscillator
model \cite{Petschulat2010}. The metaatoms are periodically arranged on a
square lattice in the $x$-$y$-plane with periods smaller than the wavelength so
that only the zeroth diffraction order propagates. A schematic of the experiment
is shown in Fig.~\ref{FIG_1}b, where we measured the transmitted intensities
through a single MM layer at normal incidence for both forward and backward
propagation direction; corresponding to positive and negative $z$-direction.

\begin{figure}
\centering
\includegraphics[width=84mm,angle=0] {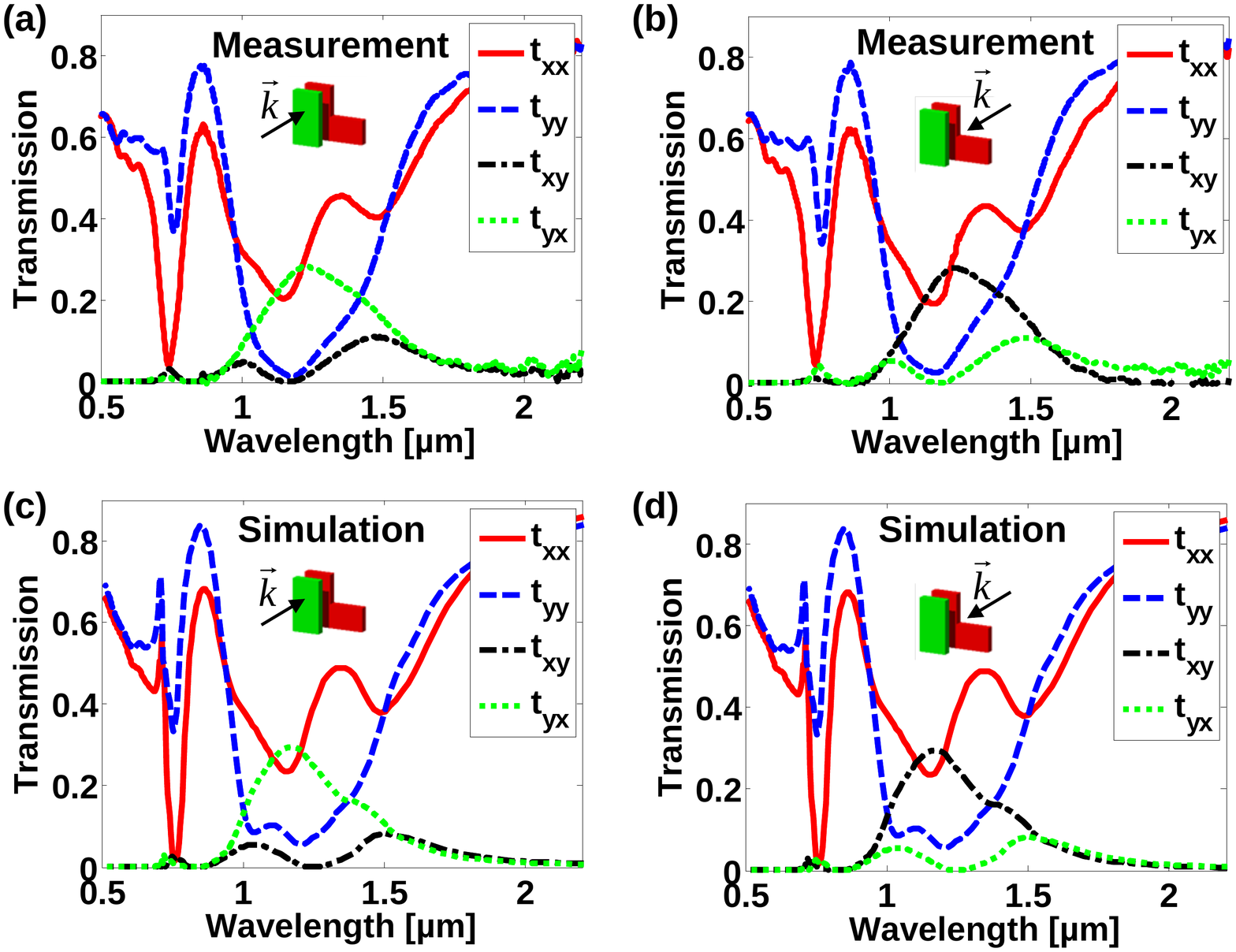}
\caption{Squared moduli of the four T-Matrix components of
the MM. a), b) Measurements and c), d) simulations. a), c) The
$k$-vector of the plane waves illuminating the MM was directed in
a), c) positive and b), d) negative $z-$direction.  Note that $t_{xy}$ and $t_{yx}$
interchange when $k$ is reversed.} \label{FIG_2}
\end{figure}

The MM was fabricated as follows: Both layers were defined in gold using
standard electron-beam lithography (Vistec SB350OS) and a lift-off technique.
Each single layer was planarized with spin-on glass resist (Dow Corning XR-1541)
matching the refractive index of the fused silica substrate. Atomic force
microscopy confirmed that the resulting surfaces were ideally planar within a
measurement accuracy of 1\,nm. For the writing of the top layer multiple
alignment marks were used to ensure an alignment accuracy of better than
20\,nm to the bottom layer (Fig.~\ref{FIG_1}c).
The squared moduli $t_{ij}=|T_{ij}|^2$ of the T-matrix of the MM were
determined experimentally (Fig.~\ref{FIG_2}a and b). Clearly $t_{xy}$ and
$t_{yx}$ interchange for opposite propagation directions proving that all
elements in the linear base are different in general. Additional deviations
between the forward and backward propagation are negligible  and can be
attributed to a non-ideal alignment of the polarizers. For comparison, the
complex transmission coefficients $T_{ij}$ are computed by using the Fourier
Modal Method \cite{Li1997} and shown in Fig.~\ref{FIG_2}c and d.
The spectral dispersion of gold was properly accounted for
\cite{Johnson1972}. The experimental and numerically computed
intensities are in almost perfect agreement. Minor deviations are
attributed to an approximative modeling of the structure with
geometrical parameters as deduced from electron
microscopy images.
As expected from the design of the metaatoms the spectra reveal a complex
multi-resonant behavior, which will be elucidated in another contribution.

\begin{figure}
\centering
\includegraphics[width=84mm,angle=0] {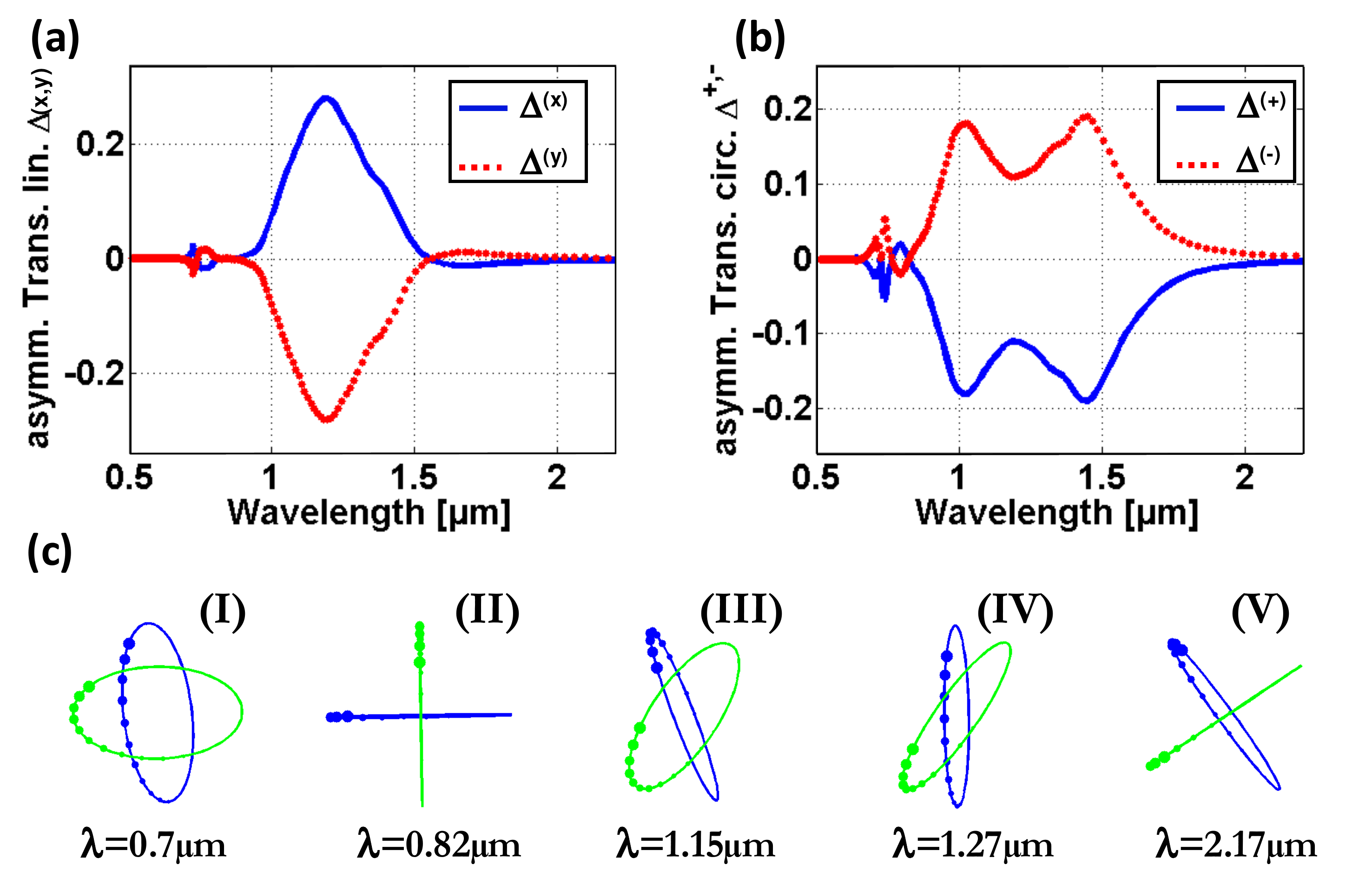}
\caption{Values for the asymmetric transmission $\Delta$ for linear
(a) and circular (b) states as determined by the numerical data. The color of the lines indicates the particular
input state. (c) Eigenstates of the polarization at different
wavelengths. The rotation direction is indicated by increasing spot sizes.} \label{FIG_3}
\end{figure}

The values for the asymmetric transmission $\Delta$ as determined from
numerical data are calculated according to Eq.~\ref{EQ_AsymmTransLin} and
Eq.~\ref{EQ_AsymmTransCirc} in the linear as well as in the circular base
(Fig.~\ref{FIG_3}a and b). Clearly, they are different for different bases and
achieve values up to 25\% in the linear base. The asymmetry for a certain state is
of course identical to the asymmetry for the complementary state with negative
sign since the total transmission for unpolarized light from both sides has to be
the same in a reciprocal medium. Figure \ref{FIG_3} represents the main result
of our work, that is an experimental prove of asymmetric transmission for
linearly polarized light in a truly three-dimensional chiral metaatom.

Any system made of complex metaatoms is characterized by its eigenstates, i.e.
those polarization states that do not change upon transmission. For these
eigenstates, being in general not orthogonal, the T-matrix is diagonal when
transformed into that eigenbase. For planar quasi-chiral structures with
additional rotational symmetries like gammadions, the eigenstates are circular
counter-rotating, whereas for anisotropic, 2D chiral structures the eigenstates
are elliptical counter-rotating with fixed phase relations. For three-dimensional
chiral systems without any additional symmetry like those considered here, the
eigenstates can be arbitrarily complex. Both elliptical co- and counter-rotating as
well as linear and circular states and combinations of them are expected. In
Fig.~\ref{FIG_3} we exemplarily show the eigenstates of the present MM for
different wavelengths. Most notably, elliptical co-rotating states (I and IV) as
well as elliptical counter-rotating states (III), almost linear states (II) and
combined ones (V) may be observed. Even the angle between the polarization
ellipses obeys no fixed relation. For structures lacking any symmetry, the
eigenstates of the polarization have obviously no preferred orientation and
rotation direction. Note that the principle form of eigenstates of the T-matrix is
invariant to any proper rotation with respect to the $z$-axis. Hence, their
complex behavior is a clear indication that it is not possible to symmetrize the
T-matrix by a proper rotation since there is no preferable orientation of the
metaatoms. We suggest that our findings can be exploited for the design of a
purely MM-based optical isolator basing on the rich variety of possible
combinations of complex T-matrices in low-symmetry and three-dimensional
chiral metaatoms.

In conclusion, we reported the first experimental observation and theoretical
analysis of asymmetric transmission for linearly polarized light in a 3D
low-symmetry MM. Contrary to previous works focusing exclusively on
circularly polarized light, the key for asymmetric transmission in any polarization
base is the complete symmetry breaking of the metaatom. The composition of
three-dimensional chiral metaatoms considerably enriches the variety of
transmission functionalities and offers yet a new freedom of design for photonic
MMs. In particular, optical isolation based on MMs seems to be in reach.

We wish to thank W. Gr\"{a}f, M. Steinert, H. Schmidt, M. Banasch,
M. Oliva and B. Steinbach for technical assistence during the sample
fabrication and S. Fahr for fruitful discussions on reciprocity.
Financial support by the Federal Ministry of Education and Research
(Unternehmen Region, ZIK ultra optics, 13N9155 and Metamat) and the
Thuringian State Government (MeMa) is acknowledged.

\end{document}